\documentclass[a4paper,fleqn]{cas-dc}

\usepackage[numbers]{natbib}
\usepackage{caption}
\usepackage{subcaption} 
\usepackage{amsthm}
\usepackage{amsmath}
\usepackage{amssymb}
\usepackage{hyperref}
\usepackage{float}

\theoremstyle{plain}

\theoremstyle{definition}

\newtheorem{remark}{Remark}

\usepackage{caption}
\usepackage{subcaption}
\usepackage{amsthm}
\usepackage{amsmath}
\usepackage{amssymb}
\usepackage{hyperref}
\usepackage{float}
\usepackage{orcidlink}  % For ORCID identifiers
\usepackage{algorithm}
\usepackage{algpseudocode}

\theoremstyle{plain}
\theoremstyle{definition}

\theoremstyle{remark}

\def\tsc#1{\csdef{#1}{\textsc{\lowercase{#1}}\xspace}}
\tsc{WGM}
\tsc{QE}
\tsc{EP}
\tsc{PMS}
\tsc{BEC}
\tsc{DE}
\begin{document}
\let\WriteBookmarks\relax
\def\floatpagepagefraction{1}
\def\textpagefraction{.001}
%\shortauthors{Farid Mafi et~al.}
\shorttitle{}
\shortauthors{}

\title [mode = title]{Bridging Data-Driven and Physics-Based Models: A Consensus Multi-Model Kalman Filter for Robust Vehicle State Estimation}                      
\tnotemark[1]

\tnotetext[1]{This document is the results of the research project funded by the Natural Sciences and Engineering Research Council of Canada (NSERC) under Grant ALLRP 566320 – 21.}

\author[1]{Farid Mafi}[type=editor,
                        auid=000,bioid=1,
                        prefix= ,
                        orcid=0000-0002-3726-0282]
\cormark[1]

\ead{farid.mafishourestani@uwaterloo.ca}

\author[1]{Ladan Khoshnevisan}[type=editor,
                        auid=000,bioid=1,
                        prefix= ,
                        orcid=0000-0002-3489-9857]
\ead{lkhoshnevisan@uwaterloo.ca}

\author[2]{Mohammad Pirani}[type=editor,
                        auid=000,bioid=1,
                        prefix= ,
                        orcid=0000-0003-2677-2140]
\ead{mpirani@uottawa.ca}

\author[1]{Amir Khajepour}[type=editor,
                        auid=000,bioid=1,
                        prefix= ,
                        orcid=0000-0002-1998-6100]  
\ead{a.khajepour@uwaterloo.ca}

\affiliation[1]{organization={Department of Mechanical and Mechatronics Engineering, University of Waterloo},
                addressline={200 University Ave W}, 
                city={Waterloo},
                postcode={N2L 3G1},
                postcodesep={},
                state={ON},
                country={Canada}}

\affiliation[2]{organization={Department of Mechanical Engineering, University of Ottawa},
                addressline={800 King Edward Ave.},
                city={Ottawa},
                postcode={K1N 6N5}, 
                postcodesep={},  
                state={ON},
                country={Canada}}

\cortext[cor1]{Corresponding author}

\begin{abstract}
Vehicle state estimation presents a fundamental challenge for autonomous driving systems, requiring both physical interpretability and the ability to capture complex nonlinear behaviors across diverse operating conditions. Traditional methodologies often rely exclusively on either physics-based or data-driven models, each with complementary strengths and limitations that become most noticeable during critical scenarios. This paper presents a novel consensus multi-model Kalman filter framework that integrates heterogeneous model types to leverage their complementary strengths while minimizing individual weaknesses. We introduce two distinct methodologies for handling covariance propagation in data-driven models: a Koopman operator-based linearization approach enabling analytical covariance propagation, and an ensemble-based method providing unified uncertainty quantification across model types without requiring pretraining. Our approach implements an iterative consensus fusion procedure that dynamically weighs different models based on their demonstrated reliability in current operating conditions. The experimental results conducted on an electric all-wheel-drive Equinox vehicle demonstrate performance improvements over single-model techniques, with particularly significant advantages during challenging maneuvers and varying road conditions, confirming the effectiveness and robustness of the proposed methodology for safety-critical autonomous driving applications.
\end{abstract}


\maketitle

\section{Introduction}

\subsection{Motivation}

Accurate vehicle state estimation is critical for autonomous driving systems, particularly during challenging scenarios like emergency maneuvers or adverse weather conditions. Traditional methods face a fundamental limitation: physics-based models offer interpretability but often fail in extreme conditions where assumptions break down, while data-driven techniques excel at capturing complex patterns but lack transparency and physical consistency. These limitations become most critical during harsh maneuvers or on low-friction surfaces—precisely when estimation accuracy is most crucial. Furthermore, autonomous systems require not just accurate estimates but also well-calibrated uncertainty quantification to navigate diverse environments safely. Our work addresses these challenges through a novel multi-model Kalman filtering framework that enables physics-based and data-driven models to collaborate in a consensus-seeking architecture, establishing a new paradigm for robust, adaptive, and uncertainty-aware vehicle state estimation.

\subsection{Literature Review}

Vehicle dynamics modeling has evolved significantly, transitioning from purely physics-based approaches to sophisticated hybrid methodologies. This evolution responds to increasing complexity of vehicle systems and demand for accurate state estimation under diverse operating conditions.

An early work in hybrid modeling came from Holzmann et al. \cite{holzmann1999vehicle}, who demonstrated one of the first applications of hybrid modeling to vehicle dynamics by combining a physical vehicle model with a neural network. Their approach showed significant error reduction when adapting to changing road conditions, though it lacked real-time implementation capabilities.

The emergence of powerful data-driven methods has revolutionized vehicle dynamics modeling. Pan et al. \cite{pan2021data} developed a data-driven approach using deep neural networks to model vehicle longitudinal dynamics, demonstrating high prediction accuracy and computational efficiency. Melzi et al. \cite{melzi2011vehicle} showed that neural networks with time-delayed inputs can effectively estimate vehicle sideslip angle without requiring a reference model or parameter identification.

Specialized neural network architectures have shown promising results for specific vehicle dynamics tasks. Wei et al. \cite{wei2016vehicle} demonstrated that General Regression Neural Network (GRNN) can effectively estimate vehicle sideslip angle by modeling it as a time series mapping of yaw speed and lateral acceleration. Bonfitto et al. \cite{bonfitto2020combined} used a multi-model approach with specialized NARX neural networks for different road conditions, selected by a classification network.

James et al. \cite{james2020longitudinal} showed that data-driven approaches outperformed traditional physical models for vehicle dynamics while requiring less computational effort for training. Devineau et al. \cite{devineau2018coupled} demonstrated that deep neural networks can effectively learn and control coupled longitudinal and lateral vehicle dynamics, outperforming conventional decoupled controllers in challenging scenarios.

Recent advancements have shown deep learning techniques outperforming traditional methods in certain conditions. Ziaukas et al. \cite{ziaukas2021estimation} showed that recurrent neural networks can effectively estimate vehicle side-slip angle across varying road surfaces without explicit friction information, outperforming sensitivity-based unscented Kalman filter methods.

Despite their performance in specific contexts, purely data-driven models often face challenges in generalization, interpretability, and uncertainty quantification. These limitations have motivated researchers to explore hybrid approaches that leverage complementary strengths of both physics-based and data-driven paradigms.

Hybrid physics-data driven modeling has emerged as a promising direction. Chen et al. \cite{chen2023hybrid} proposed a hybrid framework integrating physical models with online learning neural networks for vehicle dynamics modeling. Li et al. \cite{li2025hybrid} proposed a hybrid observer that integrates ridge regression with a dual-attention LSTM network for vehicle dynamics state estimation, demonstrating superior accuracy over traditional UKF observers.

The effectiveness of hybrid methodologies has been demonstrated across various aspects of vehicle dynamics. Chen et al. \cite{chen2024hybrid} developed a hybrid modeling approach for vehicle lateral dynamics that combines a physical model with a neural network for modeling residuals. Zhou et al. \cite{zhou2021hybrid} proposed a hybrid lateral dynamics model that uses neural networks to predict time-varying parameters for a bicycle model, demonstrating improved accuracy over conventional models in nonlinear regimes.

The integration of physical knowledge into neural networks has shown significant benefits. Robinson et al. \cite{robinson2022physics} demonstrated that injecting physical knowledge into neural networks improved prediction accuracy and reduced uncertainty for various nonlinear dynamical systems. Da et al. \cite{da2020modelling} showed that neural networks with physics-inspired pre-wired structures outperform conventional black-box networks in modeling vehicle longitudinal dynamics.

A critical distinction in hybrid approaches is between those capable of online adaptation versus those relying on offline training. Zhou et al. \cite{zhou2024vehicle} introduced a hybrid architecture that embeds neural networks into physical vehicle dynamics models to characterize modeling errors and identify time-varying parameters. Zheng et al. \cite{zheng2024koopman} demonstrated the effectiveness of a hybrid approach that integrates physics-based nominal models with data-driven Koopman models for automated vehicle control under extreme conditions.

Kalman filtering techniques have been widely applied to vehicle dynamics estimation. Rafatnia et al. \cite{rafatnia2022estimation} developed an adaptive IMU/GNSS data fusion algorithm that enhances vehicle dynamic models through complementary terms derived from sensor measurements. Chen et al. \cite{chen2022longitudinal} proposed a hierarchical vehicle dynamics state estimation approach combining adaptive square-root cubature Kalman filtering with a partitioned similarity-principle algorithm.

The challenge of varying operating conditions has motivated adaptive filtering approaches. Li et al. \cite{li2014variable} developed a Variable Structure Extended Kalman Filter that effectively combines kinematic and dynamic vehicle models. Wang et al. \cite{wang2021adaptive} developed an Adaptive Fault-Tolerant Extended Kalman Filter (AFTEKF) for vehicle state estimation that effectively handles partial sensor data loss.

Multi-model approaches, particularly Interacting Multiple Models (IMM), have gained prominence for their ability to handle systems with multiple operating modes or uncertain model structures. Tufano et al. \cite{tufano2024vehicle} developed a state-dependent Interacting Multiple Model approach for vehicle sideslip angle estimation that adapts transition probabilities between different road condition models in real-time. Wenkang et al. \cite{wenkang2021vehicle} proposed an Interacting Multiple Model based on Square Root Cubature Kalman Filter (IMM-SCKF) for vehicle state estimation.

The effectiveness of IMM approaches for handling varying road conditions has been demonstrated. Kim et al. \cite{kim2018interacting} showed that an Interacting Multiple Model Kalman filter with distinct vehicle dynamic models significantly improves lateral motion estimation compared to single-model approaches. Tsunashima et al. \cite{tsunashima2006vehicle} developed an IMM approach for simultaneously estimating vehicle side-slip angle and road friction coefficient using multiple tire models.

Recent developments in multi-model approaches have further refined these methods. Park et al. \cite{park2022vehicle} developed an IMM Kalman filter that optimally combines kinematic and dynamic vehicle models for sideslip angle estimation using low-cost sensor fusion. Cao et al. \cite{cao2023vehicle} developed a multi-physical model fusion framework using IMM and time-series analysis for vehicle sideslip trajectory prediction.

Jin et al. \cite{jin2021new} provide an analysis of hybrid-driven state estimation techniques, categorizing them into three primary architectural approaches: (1) Network-based correction of model estimates, (2) Learning of system parameters, and (3) Direct state estimation. Kim et al. \cite{kim2020vehicle} proposed combining deep ensemble neural networks with Kalman filters for vehicle sideslip angle estimation, where the neural network provides both state estimates and uncertainty values to adjust the Kalman filter's covariance matrix.

More recent advancements in neural-Kalman integration have pushed boundaries. Bertipaglia et al. \cite{bertipaglia2024unscented} proposed a mutualistic hybrid approach integrating a CNN with a UKF through end-to-end training. Nguyen et al. \cite{nguyen2023neural} developed a neural-network-based observer for vehicle dynamics that uses a sliding mode observer to identify model uncertainties from training data.

A critical aspect of effective hybrid modeling is quantification of estimation uncertainty. Lakshminarayanan et al. \cite{lakshminarayanan2017simple} demonstrated that deep ensembles provide well-calibrated uncertainty estimates that outperform Bayesian approaches while remaining simple to implement. Zhang et al. \cite{zhang2020data} developed a distributed cyber-physical system framework that integrates neural networks for behavior prediction with vehicle dynamic models for cruise control.

Based on the surveyed literature, Table~\ref{tab:trends} summarizes the key trends and research opportunities that inform our approach to vehicle dynamics modeling and state estimation.

\begin{table*}[t]
\centering
\caption{Key Trends and Research Opportunities in Vehicle Dynamics Modeling and State Estimation}
\label{tab:trends}
\small
\begin{tabular}{|p{0.27\textwidth}|p{0.53\textwidth}|p{0.12\textwidth}|}
\hline
\textbf{Research Trend} & \textbf{Description} & \textbf{References} \\
\hline
Hybrid Approach Superiority & Hybrid physics-data driven approaches consistently outperform both purely physics-based and purely data-driven methods across various vehicle dynamics applications & \cite{chen2023hybrid, li2025hybrid, robinson2022physics, zhou2021hybrid, chen2024hybrid} \\
\hline
Multi-Model Advantages & Multi-model approaches, particularly IMM frameworks, demonstrate significant advantages for handling diverse operating conditions and uncertain model structures & \cite{tufano2024vehicle, kim2018interacting, park2022vehicle, tsunashima2006vehicle, wenkang2021vehicle} \\
\hline
Neural-Kalman Integration & Integration of neural networks with Kalman filtering techniques combines adaptive learning with principled uncertainty handling & \cite{kim2020vehicle, bertipaglia2024unscented, nguyen2023neural, zhang2020data} \\
\hline
Online Adaptation & Adaptation of hybrid models to changing conditions presents significant challenges for maintaining performance across diverse operating environments & \cite{zhou2024vehicle, wang2021adaptive, li2014variable, rafatnia2022estimation} \\
\hline
Uncertainty Quantification & Meaningful confidence estimation is essential for reliable state estimation in autonomous driving systems & \cite{lakshminarayanan2017simple, bertipaglia2024unscented, kim2020vehicle} \\
\hline
Transitional Condition Management & Handling transitions between different operating regimes (e.g., normal to emergency driving) represents a key challenge for current approaches & \cite{cao2023vehicle, tufano2024vehicle, kim2018interacting} \\
\hline
\end{tabular}
\end{table*}

\subsection{Contributions}

The main contributions of this paper can be summarized as follows:

\begin{itemize}
    \item We propose a novel consensus multi-model Kalman filter framework that seamlessly integrates heterogeneous model types—both physics-based and data-driven—for robust vehicle state estimation across diverse operating conditions.

    \item We develop two distinct methodologies for uncertainty propagation in heterogeneous models: (1) a Koopman operator-based linearization approach enabling analytical covariance propagation, and (2) an ensemble-based method providing unified statistical uncertainty quantification; both integrated within an iterative consensus fusion procedure that dynamically weighs different models based on their demonstrated reliability.
    
    \item We provide a comprehensive algorithmic implementation addressing practical considerations including model error estimation, cross-model state space transformations, and formal uncertainty quantification mechanisms, while solving the challenges of model transition during regime changes to enable smooth estimation across different operating conditions.
    
    \item We demonstrate comprehensive experimental validation on an electric all-wheel-drive Equinox vehicle across diverse driving scenarios, confirming superior performance of the proposed methodology compared to single-model approaches, particularly during challenging maneuvers and varying road conditions.    
\end{itemize}

\section{A Robust Consensus Multi-Model Approach for Estimation}

Before proceeding with the detailed description of our approach, we define the key notation used throughout this paper. Table~\ref{tab:notation} provides a comprehensive summary of all symbols organized by category, including states, measurements, model parameters, covariance matrices, and operators used in our multi-model fusion framework.

\begingroup
\renewcommand{\arraystretch}{1.0}
\setlength{\tabcolsep}{4pt}
\begin{table*}[t]
\centering
\caption{Summary of notation used throughout the paper}
\label{tab:notation}
\scriptsize
\begin{tabular}{|p{0.08\textwidth}|p{0.32\textwidth}|p{0.05\textwidth}||p{0.08\textwidth}|p{0.32\textwidth}|p{0.05\textwidth}|}
\hline
\textbf{Symbol} & \textbf{Description} & \textbf{Domain} & \textbf{Symbol} & \textbf{Description} & \textbf{Domain} \\
\hline
\multicolumn{3}{|c||}{\textbf{States and Measurements}} & \multicolumn{3}{c|}{\textbf{Koopman-related Notation}} \\
\hline
$\mathbf{s}_t$ & True system state vector at time $t$ & $\mathbb{R}^n$ & $\mathbf{z}_t$ & Lifted state in Koopman space, $d > n$ & $\mathbb{R}^d$ \\
$\mathbf{x}^f_{m,t}$ & Forecast state from model $m$ at time $t$ & $\mathbb{R}^n$ & $\mathbf{A}_K$ & Linear dynamics matrix in Koopman space & $\mathbb{R}^{d \times d}$ \\
$\mathbf{x}^a_{m,t}$ & Analysis state from model $m$ at time $t$ & $\mathbb{R}^n$ & $\mathbf{B}_K$ & Control input matrix in Koopman space & $\mathbb{R}^{d \times q}$ \\
$\mathbf{y}_t$ & Measurement vector at time $t$ & $\mathbb{R}^p$ & $\mathbf{H}_K$ & Bilinear term matrix in Koopman space & $\mathbb{R}^{d \times dq}$ \\
\hline
\multicolumn{3}{|c||}{\textbf{Models and Indices}} & $\mathbf{C}_K$ & Output matrix for Koopman states & $\mathbb{R}^{n \times d}$ \\
\hline
$m$ & Model index & $\{1,\ldots,M\}$ & $\mathbf{u}_t$ & Control input vector at time $t$ & $\mathbb{R}^q$ \\
$f_m(\cdot)$ & State transition function for model $m$ & -- & $\mathbf{v}_t$ & Deterministic term in LTV system & $\mathbb{R}^d$ \\
$N_m$ & Ensemble size for model $m$ (Method 2) & $\mathbb{N}$ & $\mathbf{w}_t$ & Process noise in Koopman space & $\mathbb{R}^d$ \\
\hline
\multicolumn{3}{|c||}{\textbf{Covariance Matrices}} & \multicolumn{3}{c|}{\textbf{Fusion-related Notation}} \\
\hline
$\mathbf{P}^f_{m,t}$ & Forecast error covariance, model $m$ & $\mathbb{R}^{n \times n}$ & $\mathbf{x}^f_{1:m,t}$ & Consensus state, models 1 to $m$ & $\mathbb{R}^n$ \\
$\mathbf{P}^a_{m,t}$ & Analysis error covariance, model $m$ & $\mathbb{R}^{n \times n}$ & $\mathbf{P}^f_{1:m,t}$ & Consensus covariance, models 1 to $m$ & $\mathbb{R}^{n \times n}$ \\
$\mathbf{Q}_{m,t}$ & Process noise covariance, model $m$ & $\mathbb{R}^{n \times n}$ & $\mathbf{K}_{m,t}$ & Model incorporation gain matrix & $\mathbb{R}^{n \times n}$ \\
$\mathbf{R}_t$ & Measurement noise covariance & $\mathbb{R}^{p \times p}$ & $\mathbf{K}_t$ & Kalman gain for measurement update & $\mathbb{R}^{n \times p}$ \\
$\mathbf{Q}_K$ & Process noise covariance (Koopman) & $\mathbb{R}^{d \times d}$ & \multicolumn{3}{c|}{\textbf{Learning-related Notation}} \\
\hline
$\mathbf{R}_K$ & Measurement noise covariance (Koopman) & $\mathbb{R}^{n \times n}$ & $\mathbf{X}_d$ & Successor lifted states matrix & $\mathbb{R}^{d \times L}$ \\
\hline
\multicolumn{3}{|c||}{\textbf{System Matrices}} & $\mathbf{\tilde{X}}_d$ & Predecessor lifted states matrix & $\mathbb{R}^{d \times L}$ \\
\hline
$\mathbf{F}_{m,t}$ & Jacobian of $f_m$ at time $t$ & $\mathbb{R}^{n \times n}$ & $\mathbf{U}$ & Control inputs matrix & $\mathbb{R}^{q \times L}$ \\
$\mathbf{H}_t$ & Measurement matrix at time $t$ & $\mathbb{R}^{p \times n}$ & $\mathbf{Y}$ & Original states matrix & $\mathbb{R}^{n \times L}$ \\
$\mathbf{T}_m$ & Transformation from reference to model $m$ & $\mathbb{R}^{n \times n}$ & $L$ & Training dataset size & $\mathbb{N}$ \\
$\mathbf{I}$ & Identity matrix & $\mathbb{R}^{n \times n}$ & $\lambda_{\{A,B,H,C,Q,R\}}$ & Regularization hyperparameters & $\mathbb{R}$ \\
\hline
\multicolumn{3}{|c||}{\textbf{Ensemble Method Notation}} & \multicolumn{3}{c|}{\textbf{Operators and Error Estimation}} \\
\hline
$\boldsymbol{\eta}_i$ & Perturbation vector for ensemble member $i$ & $\mathbb{R}^n$ & $\otimes$ & Kronecker product & -- \\
$\mathbf{S}$ & Sampling covariance matrix & $\mathbb{R}^{n \times n}$ & $\odot$ & Khatri-Rao product & -- \\
$\overline{\mathbf{x}}^f_{m,t+1}$ & Ensemble mean for model $m$ & $\mathbb{R}^n$ & $\mathbb{E}[\cdot]$ & Expected value operator & -- \\
$(\mathbf{x}^f_{m,t+1})_i$ & State of $i$-th ensemble member & $\mathbb{R}^n$ & $\mathbf{d}_{m,t}$ & Innovation vector for model $m$ & $\mathbb{R}^p$ \\
\hline
\end{tabular}
\end{table*}
\endgroup

\subsection{Preliminaries: Multi-Model Ensemble Kalman Filter}

In many real-world applications, multiple prediction models of different types—both data-driven and physics-based dynamic models—may be available for state estimation of a complex system. The models may have complementary strengths and weaknesses, and combining them intelligently can yield more accurate and robust predictions than using any single model alone. We present a consensus-based multi-model approach that can effectively integrate heterogeneous models while respecting their unique characteristics.

Consider a system with state vector $\mathbf{s}_t \in \mathbb{R}^n$ at time $t$. Suppose we have $M$ different models (indexed by $m = 1, 2, \ldots, M$) that provide state predictions, where some models are physics-based dynamic models and others are data-driven models such as neural networks or kernel-based approaches. Each model $m$ provides a prediction of the state at time $t$, denoted by $\mathbf{x}^f_{m,t}$, with an associated error covariance matrix $\mathbf{P}^f_{m,t}$.

The multi-model ensemble Kalman filter framework proceeds as follows:
\begin{enumerate}
    \item Each model produces its state prediction $\mathbf{x}^f_{m,t}$ and associated error covariance $\mathbf{P}^f_{m,t}$.
    \item These predictions are combined to form a consensus state estimate before observation data is incorporated.
    \item The consensus prediction is then updated using the observation data to produce the final state estimate.
\end{enumerate}

The challenge in this framework lies in obtaining reliable error covariance estimates for data-driven models, which typically do not provide them directly. We present two methodologies that address this challenge differently.

\subsection{Novel Extensions for Heterogeneous Model Integration}

We propose two distinct methodologies for handling the integration of data-driven and dynamic models within the MM-EnKF framework. Both methods differ primarily in their prediction steps and covariance updates. The choice of each method, along with its advantages and limitations, is discussed in detail in Subsection~\ref{subsec:implementation}.

\subsubsection{Method 1: Koopman Linearization for Data-Driven Models}

Our first methodology leverages the Koopman operator theory to linearize data-driven models, enabling consistent covariance propagation within the Kalman filter framework. The key insight is to lift the original nonlinear system into a higher-dimensional Reproducing Kernel Hilbert Space (RKHS), where the system becomes bilinear and amenable to analytical treatment.

\paragraph{Prediction Step for Data-Driven Models:}

For a data-driven model like a neural network, the state prediction is directly given by the model output:
\begin{align}
    \mathbf{x}^f_{m,t+1} = f_m(\mathbf{x}^a_{m,t})
\end{align}
where $f_m$ represents the data-driven model $m$, and $\mathbf{x}^a_{m,t}$ is the analysis state at time $t$. The training data for these models should consist of time series observations to capture the temporal dynamics of the system.

\paragraph{Koopman Linearization for Covariance Propagation:}

While the data-driven model directly provides the state prediction, it does not naturally provide a way to propagate the error covariance. This is where the Koopman linearization becomes crucial. Following the approach presented in \cite{guo2021koopman}, we lift the original system into a higher-dimensional space where it has a bilinear representation.

For a control-affine system, we can express the lifted dynamics as:
\begin{align}
    \mathbf{z}_{t+1} = \mathbf{A}_K\mathbf{z}_t + \mathbf{B}_K\mathbf{u}_t + \mathbf{H}_K(\mathbf{u}_t \otimes \mathbf{z}_t) + \mathbf{w}_t
     \label{eq:z_dynamics}
\end{align}
where $\mathbf{z}_t$ is the lifted state in the Koopman space at time $t$, $\mathbf{u}_t$ is the control input, $\mathbf{w}_t$ is the process noise, and $\mathbf{A}_K$, $\mathbf{B}_K$, and $\mathbf{H}_K$ are the system matrices in the lifted space. The symbol $\otimes$ represents the Kronecker product.

The system matrices are learned by solving a least-squares problem on training trajectories, with a loss function $V = V_1 + V_2$ where:\begin{equation}
\begin{aligned}
V_1 &= \frac{1}{2}\|\mathbf{X}_d - \mathbf{A}_K\mathbf{\tilde{X}}_d - \mathbf{B}_K\mathbf{U} - \mathbf{H}_K(\mathbf{U} \odot \mathbf{\tilde{X}}_d)\|^2_{\mathbf{Q}_K^{-1}} \\
&\quad + \frac{1}{2}\|\mathbf{Y} - \mathbf{C}_K\mathbf{X}_d\|^2_{\mathbf{R}_K^{-1}} - \frac{1}{2}L\ln|\mathbf{Q}_K^{-1}| - \frac{1}{2}L\ln|\mathbf{R}_K^{-1}|
\end{aligned}
\end{equation}

\begin{equation}
\begin{aligned}
V_2 &= \frac{1}{2}L\lambda_A\|\mathbf{A}_K\|^2_{\mathbf{Q}_K^{-1}} + \frac{1}{2}L\lambda_B\|\mathbf{B}_K\|^2_{\mathbf{Q}_K^{-1}} \\
&\quad + \frac{1}{2}L\lambda_H\|\mathbf{H}_K\|^2_{\mathbf{Q}_K^{-1}} + \frac{1}{2}L\lambda_C\|\mathbf{C}_K\|^2_{\mathbf{R}_K^{-1}} \\
&\quad + \frac{1}{2}L\lambda_Q\text{tr}(\mathbf{Q}_K^{-1}) + \frac{1}{2}L\lambda_R\text{tr}(\mathbf{R}_K^{-1})
\end{aligned}
\end{equation}

Here, $V_1$ represents the negative log-likelihood of fitting the data, and $V_2$ includes regularization terms with hyperparameters $\lambda_A, \lambda_B, \lambda_H, \lambda_C, \lambda_Q, \lambda_R$. The matrices and parameters appearing in these equations follow the definitions provided in Table~\ref{tab:notation}, with $\odot$ representing the Khatri-Rao product.

During the testing phase—i.e., when evaluating the system under previously unseen scenarios and given the control inputs— the bilinear system \eqref{eq:z_dynamics} can be algebraically manipulated into a linear time-varying (LTV) system:
\begin{align}
    \mathbf{z}_{t+1} = \mathbf{A}_{K,t}\mathbf{z}_t + \mathbf{v}_t + \mathbf{w}_t
\end{align}
where $\mathbf{A}_{K,t} = \mathbf{A}_K + \mathbf{H}_K(\mathbf{u}_t \otimes \mathbf{I})$ and $\mathbf{v}_t = \mathbf{B}_K\mathbf{u}_t$.

This LTV representation enables us to update the covariance matrix using standard linear Kalman filter equations:
\begin{align}
    \mathbf{P}^f_{m,t+1} = \mathbf{A}_{K,t}\mathbf{P}^a_{m,t}\mathbf{A}_{K,t}^T + \mathbf{Q}_{m,t}
\end{align}
where $\mathbf{P}^a_{m,t}$ is the analysis error covariance at time $t$, and $\mathbf{Q}_{m,t}$ is the process noise covariance matrix for model $m$ at time $t$. 

For traditional dynamic models, the standard prediction step is used:
\begin{align}
    \mathbf{x}^f_{m,t+1} &= f_m(\mathbf{x}^a_{m,t})\\
    \mathbf{P}^f_{m,t+1} &= \mathbf{F}_{m,t}\mathbf{P}^a_{m,t}\mathbf{F}_{m,t}^T + \mathbf{Q}_{m,t}
\end{align}
where $\mathbf{F}_{m,t}$ is the Jacobian of $f_m$ evaluated at $\mathbf{x}^a_{m,t}$.

The key advantage of Method 1 is that it adapts data-driven models to the framework established for dynamic models, allowing for a more straightforward integration within existing Kalman filter implementations.

\begin{remark}
The Koopman linearization approach follows a fundamental principle similar to other techniques in machine learning and dynamical systems: transforming complex nonlinear problems into more tractable linear ones through embedding in higher-dimensional spaces. This concept closely parallels kernel methods in Support Vector Machines (SVMs), where nonlinearly separable data is mapped into higher-dimensional feature spaces via kernel functions (e.g., Radial Basis Function, polynomial kernels) to enable linear separation. Both approaches leverage the insight that while a system may exhibit highly nonlinear behavior in its original state space, there often exists a higher-dimensional representation where the dynamics become approximately linear. This principle of "lifting" to achieve linearization appears across various disciplines, including reproducing kernel Hilbert spaces (RKHS) methods, feature engineering in machine learning, and embedding theorems in dynamical systems theory. The Koopman approach specifically exploits this principle for the propagation of uncertainty in nonlinear dynamical systems.
\end{remark}

\subsubsection{Method 2: Ensemble-Based Covariance Propagation}

Our second methodology is inspired by ensemble Kalman filter techniques \cite{bach2023multi} and treats data-driven models similarly to dynamic models from a statistical perspective. This method does not require pretraining a separate linearization model and is particularly suitable when online adaptability is desired.

\paragraph{Ensemble Generation}

For each model $m$, we generate an ensemble of $N_m$ perturbed state predictions by sampling from a normal distribution around the model's prediction:\begin{equation}
\begin{aligned}
    (\mathbf{x}^f_{m,t+1})_i &= f_m(\mathbf{x}^a_{m,t}) + \boldsymbol{\eta}_i, \\
    \text{where } \boldsymbol{\eta}_i &\sim \mathcal{N}(\mathbf{0}, \mathbf{S}), \quad i=1,\ldots,N_m
\end{aligned}
\end{equation}
and $\mathbf{S}$ is an appropriate sampling covariance matrix, which may be based on historical model errors or set to a suitable initial value and adapted over time. The parameter $N_m$ represents the ensemble size for model $m$.

\paragraph{Covariance Estimation}

\begin{figure}
    \centering
    \includegraphics[width=0.95\columnwidth]{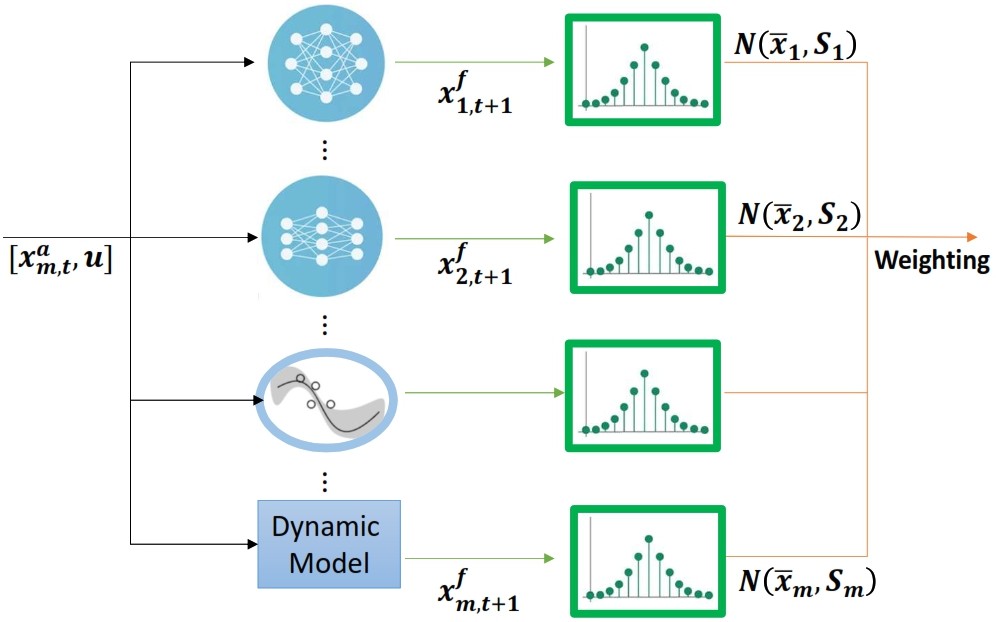}
    \caption{Schematic of Method 2: Ensemble-Based Covariance Propagation. Both data-driven and dynamic models use the same framework for covariance estimation based on ensemble perturbation.}
    \label{fig:method2}
\end{figure}

The forecast error covariance is then estimated directly from the ensemble spread:
\begin{equation}
\begin{aligned}
    \mathbf{P}^{f}_{m,t+1} = \frac{1}{N_m-1}\sum_{i=1}^{N_m}(&(\mathbf{x}^f_{m,t+1})_i - \overline{\mathbf{x}}^f_{m,t+1})\\
    &\times((\mathbf{x}^f_{m,t+1})_i - \overline{\mathbf{x}}^f_{m,t+1})^T
\end{aligned}
\end{equation}
where $\overline{\mathbf{x}}^f_{m,t+1}$ is the ensemble mean, which serves as the model's prediction. Figure~\ref{fig:method2} illustrates this ensemble-based approach to covariance propagation.

This technique has the advantage of treating data-driven and dynamic models in a unified framework, as the same ensemble-based covariance estimation can be applied to both types of models. Dynamic models simply use their own state transition functions instead of neural networks or other data-driven approaches.

\subsection{Consensus Multi-Model Fusion}

Once the predictions and their associated covariances from all models are obtained, we proceed with the consensus fusion step, which is identical for both the Koopman-based and Ensemble-based methods as previously described. This step combines the predictions from different models based on their relative uncertainties before incorporating observation data.

\subsubsection{Iterative Fusion Procedure}

We adopt an iterative approach to fuse the model predictions. Starting with the first model's prediction as the initial consensus, we sequentially incorporate each additional model. By Initializing with the first model:
\begin{align}
    \mathbf{x}^{f}_{1:1,t} &= \mathbf{x}^f_{1,t} \\
    \mathbf{P}^{f}_{1:1,t} &= \mathbf{P}^f_{1,t}
\end{align}
For each subsequent model $m = 2, \ldots, M$, compute:
\begin{align}
    \mathbf{K}_{m,t} &= \mathbf{P}^{f}_{1:m-1,t}\mathbf{T}_m^T(\mathbf{T}_m\mathbf{P}^{f}_{1:m-1,t}\mathbf{T}_m^T + \mathbf{P}^f_{m,t})^{-1} \\
    \mathbf{x}^{f}_{1:m,t} &= \mathbf{x}^{f}_{1:m-1,t} + \mathbf{K}_{m,t}(\mathbf{x}^f_{m,t} - \mathbf{T}_m\mathbf{x}^{f}_{1:m-1,t}) \\
    \mathbf{P}^{f}_{1:m,t} &= (\mathbf{I} - \mathbf{K}_{m,t}\mathbf{T}_m)\mathbf{P}^{f}_{1:m-1,t}
\end{align}

Here, $\mathbf{T}_m$ is a transformation operator that maps from the reference model space to the space of model $m$. For models operating in the same state space, $\mathbf{T}_m$ is simply the identity matrix. The notation $\mathbf{x}^{f}_{1:m,t}$ represents the consensus forecast state combining models 1 through $m$ at time $t$. After incorporating all models, we obtain the consensus prediction $\mathbf{x}^{f}_{1:M,t}$ with covariance $\mathbf{P}^{f}_{1:M,t}$.

\subsubsection{Observation Update}

Finally, we update the consensus prediction using the observation data:
\begin{align}
    \mathbf{K}_t &= \mathbf{P}^{f}_{1:M,t}\mathbf{H}_t^T(\mathbf{H}_t\mathbf{P}^{f}_{1:M,t}\mathbf{H}_t^T + \mathbf{R}_t)^{-1} \\
    \mathbf{x}^{a}_{t} &= \mathbf{x}^{f}_{1:M,t} + \mathbf{K}_t(\mathbf{y}_{t} - \mathbf{H}_t\mathbf{x}^{f}_{1:M,t}) \\
    \mathbf{P}^{a}_{t} &= (\mathbf{I} - \mathbf{K}_t\mathbf{H}_t)\mathbf{P}^{f}_{1:M,t}
\end{align}
where $\mathbf{y}_{t}$ is the observation vector at time $t$, $\mathbf{H}_t$ is the observation matrix at time $t$, and $\mathbf{R}_t$ is the observation error covariance matrix at time $t$.

\subsection{Implementation and Method Comparison}
\label{subsec:implementation}

Our framework offers two complementary approaches to covariance propagation for data-driven models, each with distinct advantages. Method 1 employs Koopman operator theory to find linearized representations of nonlinear data-driven models, enabling analytical covariance propagation. While requiring offline training, this approach is particularly effective when the system has known structural properties to guide the lifting process. Conversely, Method 2 offers a more straightforward approach by treating all models within a unified ensemble framework. It avoids pretraining requirements and adapts naturally as new data becomes available, though it may demand larger ensemble sizes for reliable covariance estimates, increasing computational demands.

When implementing either approach, several critical considerations emerge. The model set should capture diverse aspects of system dynamics without redundancy, while computational resources must be carefully balanced—selecting appropriate lifted space dimensions for Method 1 or ensemble sizes for Method 2. The framework's effectiveness depends on proper state space transformations when models use different representations, and for Method 2, initialization of sampling covariance based on prior knowledge or historical errors is essential. Additionally, adaptive weighting of models based on their demonstrated prediction accuracy enables dynamic adjustment to different operating regimes.

A fundamental component across both methods is proper model error estimation. We employ innovation statistics to estimate model error covariances:
\begin{align}
    \mathbf{d}_{m,t} &= \mathbf{y}_t - \mathbf{H}_t\mathbf{x}^f_{m,t}\\
    \mathbb{E}[\mathbf{d}_{m,t}\mathbf{d}_{m,t}^T] &= \mathbf{H}_t\mathbf{P}^f_{m,t}\mathbf{H}_t^T + \mathbf{R}_t
\end{align}

This innovation-based approach enables propagation of error statistics back to the model space, ensuring appropriate calibration of relative uncertainties between different model types—a critical factor for the effective fusion of heterogeneous models.

\subsection{Algorithmic Summary}

Algorithm~\ref{alg:mmenkf} presents a comprehensive implementation of our proposed Multi-Model Ensemble Kalman Filter framework with heterogeneous model integration. This algorithm encapsulates the entire workflow described in the previous sections, including both covariance propagation methods, the iterative consensus fusion procedure, and the observation update step. The algorithm demonstrates how our approach seamlessly integrates data-driven and physics-based models within a unified estimation framework while accounting for their unique characteristics in uncertainty propagation. By following this algorithmic implementation, practitioners can effectively combine the complementary strengths of diverse modeling paradigms for enhanced state estimation performance across varying operating conditions.

\begin{algorithm*}[t]
\caption{Multi-Model Ensemble Kalman Filter with Heterogeneous Model Integration}
\label{alg:mmenkf}
\begin{algorithmic}[1]
\Statex \textbf{Initialization}
\State Initialize analysis state $\mathbf{x}^a_{m,0}$ and covariance $\mathbf{P}^a_{m,0}$ for each model $m = 1, 2, \ldots, M$

\Statex
\For{each time step $t = 0, 1, 2, \ldots$}
    \Statex \textbf{1. Model-specific forecasting}
    \For{each model $m = 1, 2, \ldots, M$}
        \If{Method 1 is used}
            \If{model $m$ is data-driven}
                \State Predict state: $\mathbf{x}^f_{m,t+1} = f_m(\mathbf{x}^a_{m,t})$
                \State Derive Koopman linearization $\mathbf{A}_{K,t}$
                \State Update covariance: $\mathbf{P}^f_{m,t+1} = \mathbf{A}_{K,t}\mathbf{P}^a_{m,t}\mathbf{A}_{K,t}^T + \mathbf{Q}_{m,t}$
            \Else \Comment{Physics-based model with Method 1}
                \State Predict state: $\mathbf{x}^f_{m,t+1} = f_m(\mathbf{x}^a_{m,t})$
                \State Compute Jacobian $\mathbf{F}_{m,t}$
                \State Update covariance: $\mathbf{P}^f_{m,t+1} = \mathbf{F}_{m,t}\mathbf{P}^a_{m,t}\mathbf{F}_{m,t}^T + \mathbf{Q}_{m,t}$
            \EndIf
        \Else \Comment{Method 2 is used (Ensemble-based approach)}
            \State Generate ensemble: $\{(\mathbf{x}^a_{m,t})_i\}_{i=1}^{N_m}$ where $(\mathbf{x}^a_{m,t})_i = \mathbf{x}^a_{m,t} + \boldsymbol{\eta}_i$
            \State Propagate each ensemble member: $(\mathbf{x}^f_{m,t+1})_i = f_m((\mathbf{x}^a_{m,t})_i)$ for $i = 1, 2, \ldots, N_m$
            \State Compute ensemble mean: $\mathbf{x}^f_{m,t+1} = \frac{1}{N_m}\sum_{i=1}^{N_m} (\mathbf{x}^f_{m,t+1})_i$
            \State Compute ensemble covariance: 
            \Statex \hspace{1.4cm} $\mathbf{P}^f_{m,t+1} = \frac{1}{N_m-1}\sum_{i=1}^{N_m}((\mathbf{x}^f_{m,t+1})_i - \mathbf{x}^f_{m,t+1})((\mathbf{x}^f_{m,t+1})_i - \mathbf{x}^f_{m,t+1})^T$
        \EndIf
    \EndFor
    
    \Statex \textbf{2. Consensus fusion of model predictions}
    \State Initialize consensus with first model: $\mathbf{x}^{f}_{1:1,t+1} = \mathbf{x}^f_{1,t+1}$, $\mathbf{P}^{f}_{1:1,t+1} = \mathbf{P}^f_{1,t+1}$
    
    \For{each additional model $m = 2, 3, \ldots, M$}
        \State Compute incorporation gain: 
        \Statex \hspace{1.4cm} $\mathbf{K}_{m,t+1} = \mathbf{P}^{f}_{1:m-1,t+1}\mathbf{T}_m^T(\mathbf{T}_m\mathbf{P}^{f}_{1:m-1,t+1}\mathbf{T}_m^T + \mathbf{P}^f_{m,t+1})^{-1}$
        \State Update consensus state: 
        \Statex \hspace{1.4cm} $\mathbf{x}^{f}_{1:m,t+1} = \mathbf{x}^{f}_{1:m-1,t+1} + \mathbf{K}_{m,t+1}(\mathbf{x}^f_{m,t+1} - \mathbf{T}_m\mathbf{x}^{f}_{1:m-1,t+1})$
        \State Update consensus covariance: 
        \Statex \hspace{1.4cm} $\mathbf{P}^{f}_{1:m,t+1} = (\mathbf{I} - \mathbf{K}_{m,t+1}\mathbf{T}_m)\mathbf{P}^{f}_{1:m-1,t+1}$
    \EndFor

    \Statex \textbf{3. Measurement update}
    \State Obtain measurement $\mathbf{y}_{t+1}$
    \State Compute Kalman gain: 
    \Statex \hspace{1.4cm} $\mathbf{K}_{t+1} = \mathbf{P}^{f}_{1:M,t+1}\mathbf{H}_{t+1}^T(\mathbf{H}_{t+1}\mathbf{P}^{f}_{1:M,t+1}\mathbf{H}_{t+1}^T + \mathbf{R}_{t+1})^{-1}$
    \State Update state with measurement: 
    \Statex \hspace{1.4cm} $\mathbf{x}^a_{t+1} = \mathbf{x}^{f}_{1:M,t+1} + \mathbf{K}_{t+1}(\mathbf{y}_{t+1} - \mathbf{H}_{t+1}\mathbf{x}^{f}_{1:M,t+1})$
    \State Update covariance: 
    \Statex \hspace{1.4cm} $\mathbf{P}^a_{t+1} = (\mathbf{I} - \mathbf{K}_{t+1}\mathbf{H}_{t+1})\mathbf{P}^{f}_{1:M,t+1}$
    
    \Statex \textbf{4. Feedback to individual models}
    \For{each model $m = 1, 2, \ldots, M$}
        \State Map consensus state to model space: $\mathbf{x}^a_{m,t+1} = \mathbf{T}_m\mathbf{x}^a_{t+1}$
        \State Map consensus covariance to model space: $\mathbf{P}^a_{m,t+1} = \mathbf{T}_m\mathbf{P}^a_{t+1}\mathbf{T}_m^T$
    \EndFor
\EndFor
\end{algorithmic}
\end{algorithm*}

\subsection{Discussion}

Our consensus multi-model framework addresses the fundamental challenge of integrating heterogeneous model types for robust state estimation. By solving the covariance propagation problem for data-driven models through either Koopman linearization or ensemble-based methods, we enable principled fusion of complementary modeling paradigms. The key benefits of our framework include:
\begin{itemize}
    \item \textbf{Robustness through complementarity}: By combining multiple models with different strengths, the framework mitigates individual model weaknesses. When certain models perform poorly in specific regimes, others can compensate, providing consistent performance across diverse operating conditions.
    
    \item \textbf{Adaptive weighting}: The covariance-based fusion mechanism naturally adjusts the influence of each model based on its demonstrated reliability in current conditions, enabling automatic adaptation to changing environments.
    
    \item \textbf{Principled uncertainty quantification}: The framework provides well-calibrated uncertainty estimates through rigorous covariance propagation, essential for risk-aware decision-making in safety-critical applications.
\end{itemize}

The choice between Methods 1 and 2 depends on application requirements. Method 1 (Koopman linearization) offers stronger analytical guarantees when the system structure permits effective linearization, while Method 2 (ensemble-based approach) provides greater flexibility and simpler implementation at the cost of higher computational demands. Method 1 requires offline training but preserves model structure, whereas Method 2 adapts more readily to changing conditions without pretraining.

The iterative consensus fusion procedure represents a key innovation, enabling incremental incorporation of diverse models while respecting their distinctive characteristics. This methodology is particularly valuable for autonomous vehicle applications, where state estimation must remain reliable across rapidly changing conditions and during critical maneuvers when traditional models often fail.

Overall, our framework establishes a new paradigm for vehicle state estimation that leverages the full spectrum of available modeling tools. By bridging physics-based and data-driven paradigms through a principled statistical framework, we provide a pathway toward more robust, adaptive, and uncertainty-aware estimation for complex dynamical systems.

\section{Experiments}
\subsection{Experimental Settings}
\begin{figure}[t]
    \centering
    \includegraphics[width=0.7\columnwidth]{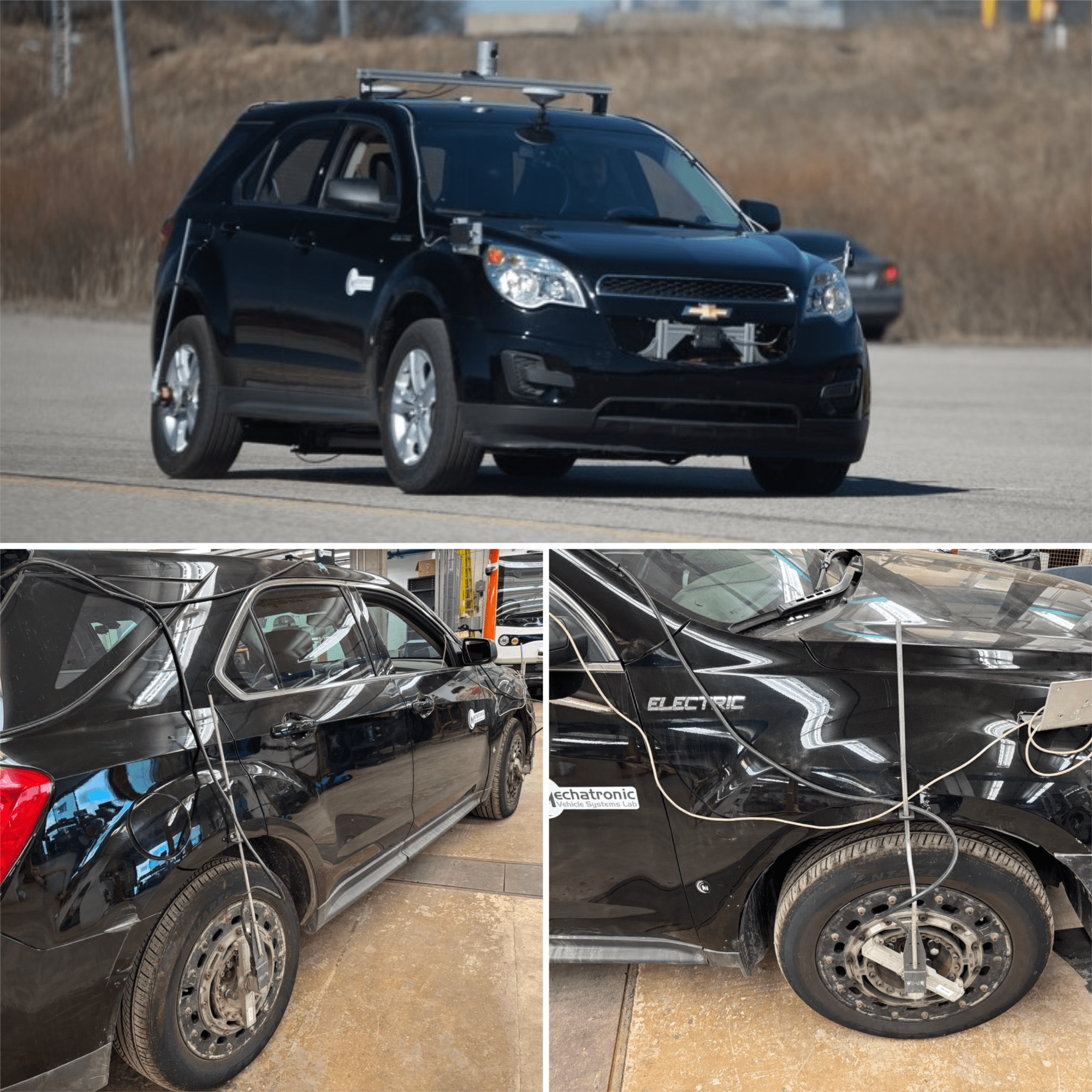}
    \caption{Electric all-wheel-drive Equinox test vehicle used for experimental validation.}
    \label{fig:test_vehicle}
\end{figure}

The proposed consensus multi-model Kalman filter framework was evaluated using an electric all-wheel-drive Chevrolet Equinox vehicle, as shown in Figure~\ref{fig:test_vehicle}. This testbed provided an ideal platform for validating our approach under diverse operating conditions. The specifications of this test vehicle are presented in Table~\ref{tab:vehicle_specs}.
\begin{table}[t]
\centering
\caption{The specifications of the test vehicle}
\label{tab:vehicle_specs}
\begin{tabular}{lc}
\hline
Parameter & Value \\
\hline
Vehicle mass ($M$) & 2271 kg \\
Vehicle moment of inertia ($I_z$) & 4600 $\text{kgm}^2$ \\
Distance from front axle to CG ($a$) & 1.42 m \\
Distance from rear axle to CG ($b$) & 1.43 m \\
Tire effective radius ($R_w$) & 0.347 m \\
Tire cornering stiffness ($C_\alpha$) & 83700 N \\
Wheel moment of inertia ($I_w$) & 1.7 $\text{kgm}^2$ \\
Steering ratio (SR) & 18:1 \\
\hline
\end{tabular}
\end{table}

The test vehicle is equipped with a robotic steering system. The vehicle's powertrain system has an electric motor for each wheel, providing independent control over each wheel's torque. The sensing suite includes a 6-axis IMU sensor that measures yaw rate and longitudinal and lateral accelerations. A 6-axis GPS sensor (RT2500) provides the vehicle's longitudinal and lateral velocities. The vehicle is also equipped with a steering wheel sensor, wheel speed sensors, and wheel torque sensors.

During each maneuver, data from the Inertial Measurement Unit (IMU), Global-Positioning-System (GPS), and wheel sensors were collected. The signal measurements were sampled at a frequency of 200 Hz for all maneuvers. To rigorously evaluate the algorithm's performance in challenging conditions, extensive testing was conducted in extreme winter environments with temperatures as low as -28°C on ice and snow surfaces, in addition to standard tests on dry and wet surfaces. The test protocol included a diverse range of demanding maneuvers: vehicle launches on varying friction surfaces, sustained acceleration runs, pulsed torque inputs, high sideslip scenarios, low-speed turning tests in both directions, and high-speed sine-steer maneuvers at 70 km/h. Model Predictive Control (MPC) was implemented with the estimation algorithms in the control loop, with various traction control settings tested to intentionally induce instability and push the vehicle beyond its linear handling regime. This comprehensive experimental design created scenarios where traditional single-model approaches would typically struggle, providing an ideal testbed for evaluating the robustness of the consensus multi-model framework under real-world challenging conditions.

\subsection{State Estimation Results}
\begin{figure}
    \centering
    \includegraphics[width=0.99\columnwidth]{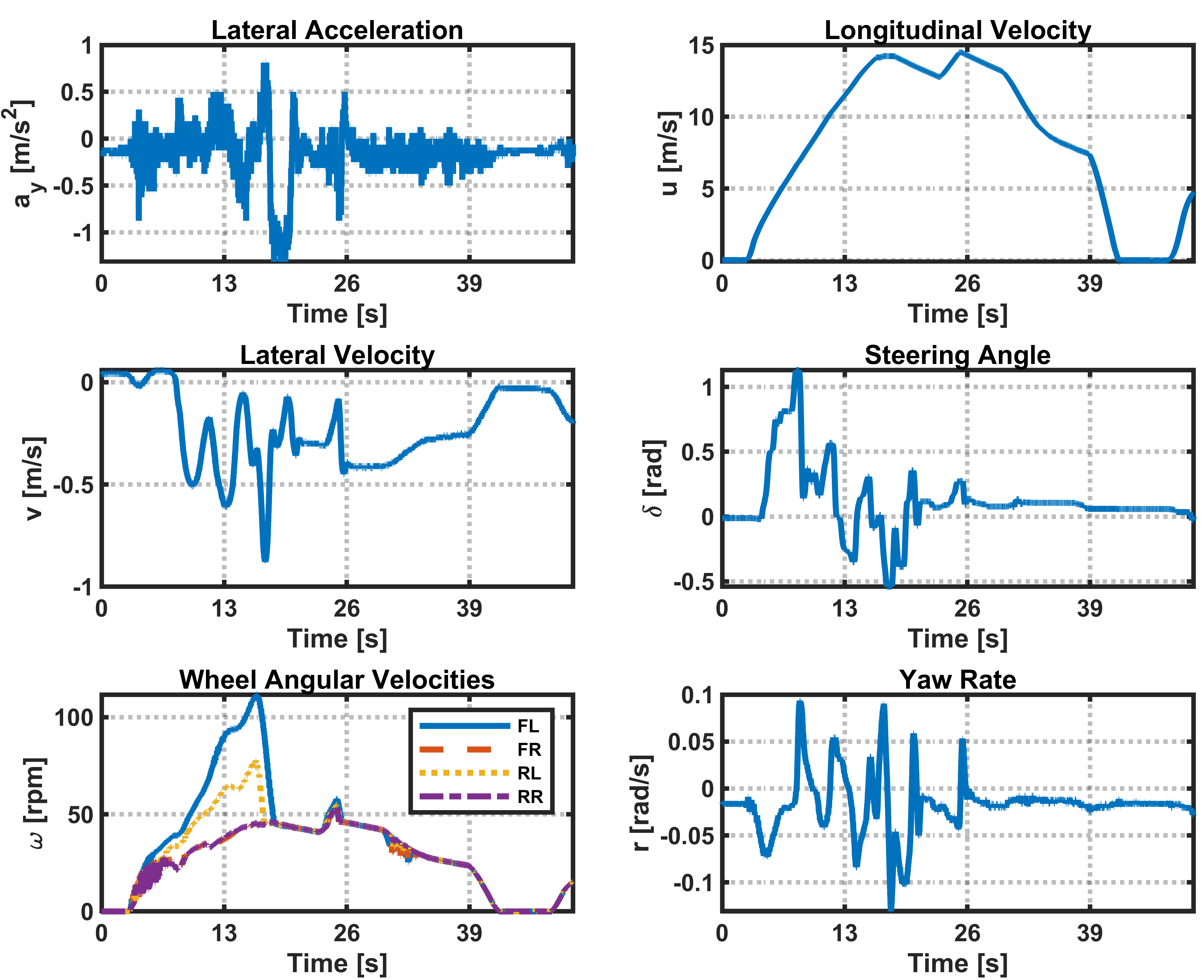}
    \caption{Ground truth data from the unseen test maneuver. This dataset serves as the benchmark for evaluating estimation performance.}
    \label{fig:Ground_truth}
\end{figure}

In this section, we present and analyze the performance of our two methodologies using experimental data collected from the test vehicle. The test dataset consists of 19 distinct maneuvers, with 18 maneuvers used for training the neural networks and one unseen maneuver reserved for validation and performance evaluation. Figure~\ref{fig:Ground_truth} depicts the ground truth data from this unseen test maneuver, which provides the benchmark for assessing estimation accuracy.

For the first methodology, the Koopman Linearization for Data-Driven Models, we implemented two neural networks(referred to as Net 1 and Net 2) and divided the time series data into equal-sized batches for training. The neural network architecture employs a dual feedforward configuration with a three-layer structure. Each network's input layer contains 8 neurons, capturing critical vehicle states including lateral and longitudinal velocities, steering angle, yaw rate, and wheel speeds. Two hidden layers with 25 and 15 neurons respectively provide sufficient capacity to model nonlinear vehicle dynamics without overfitting. The output layer consists of 6 neurons that predict state variables for the subsequent time step.

Both networks were trained using the \textit{Levenberg} backpropagation algorithm, which offers superior convergence properties for medium-sized networks compared to traditional gradient descent methods. The networks were trained independently on distinct datasets with normalized inputs and outputs, enhancing generalization capabilities while maintaining numerical stability. A critical aspect of our approach is that unmeasurable states like lateral velocity are only used during network training with observed data. During testing with unseen data, these states are treated as completely unobservable, which addresses the primary challenge of state estimation in vehicle dynamics applications.

\begin{figure}
    \centering
    \includegraphics[width=0.95\columnwidth]{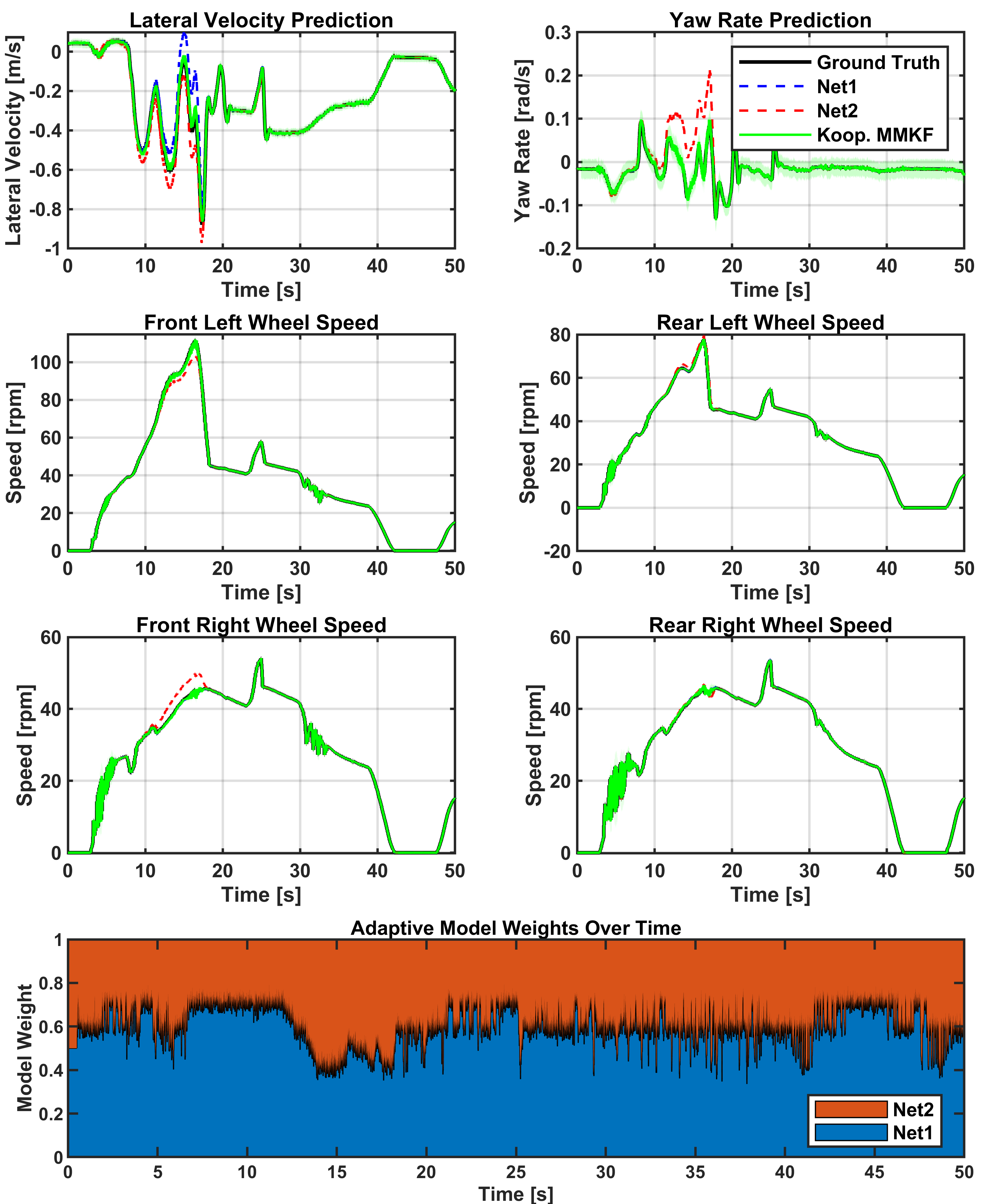}
    \caption{Performance of Method 1: Koopman Linearization for Data-Driven Models.}
    \label{fig:Koopman_results}
\end{figure}

The implementation further leverages Koopman operator theory to provide a principled framework for covariance propagation within the Multi-Model Kalman Filter. We employed Random Fourier Features (RFF) as lifting functions to transform the state and control inputs into an expanded observable space, where bilinear dynamics are converted to linear time-varying (LTV) systems through strategic tensor products. To enhance stability, we implemented a blending technique between the Koopman-predicted lifted state and direct state embedding with parameter $\alpha=0.8$, effectively mitigating potential divergence issues. The covariance matrices underwent careful conditioning procedures to ensure numerical stability and positive-definiteness throughout the estimation process. The performance results of this method are presented in Figure~\ref{fig:Koopman_results}, which demonstrates excellent estimation accuracy. This hybrid methodology—utilizing neural networks for state prediction while leveraging Koopman operators for covariance propagation—represents a sophisticated fusion of data-driven prediction with theoretically grounded uncertainty propagation, particularly advantageous for estimating unmeasurable states like lateral velocity in complex vehicle dynamics applications.

\begin{figure}
    \centering
    \includegraphics[width=0.95\columnwidth]{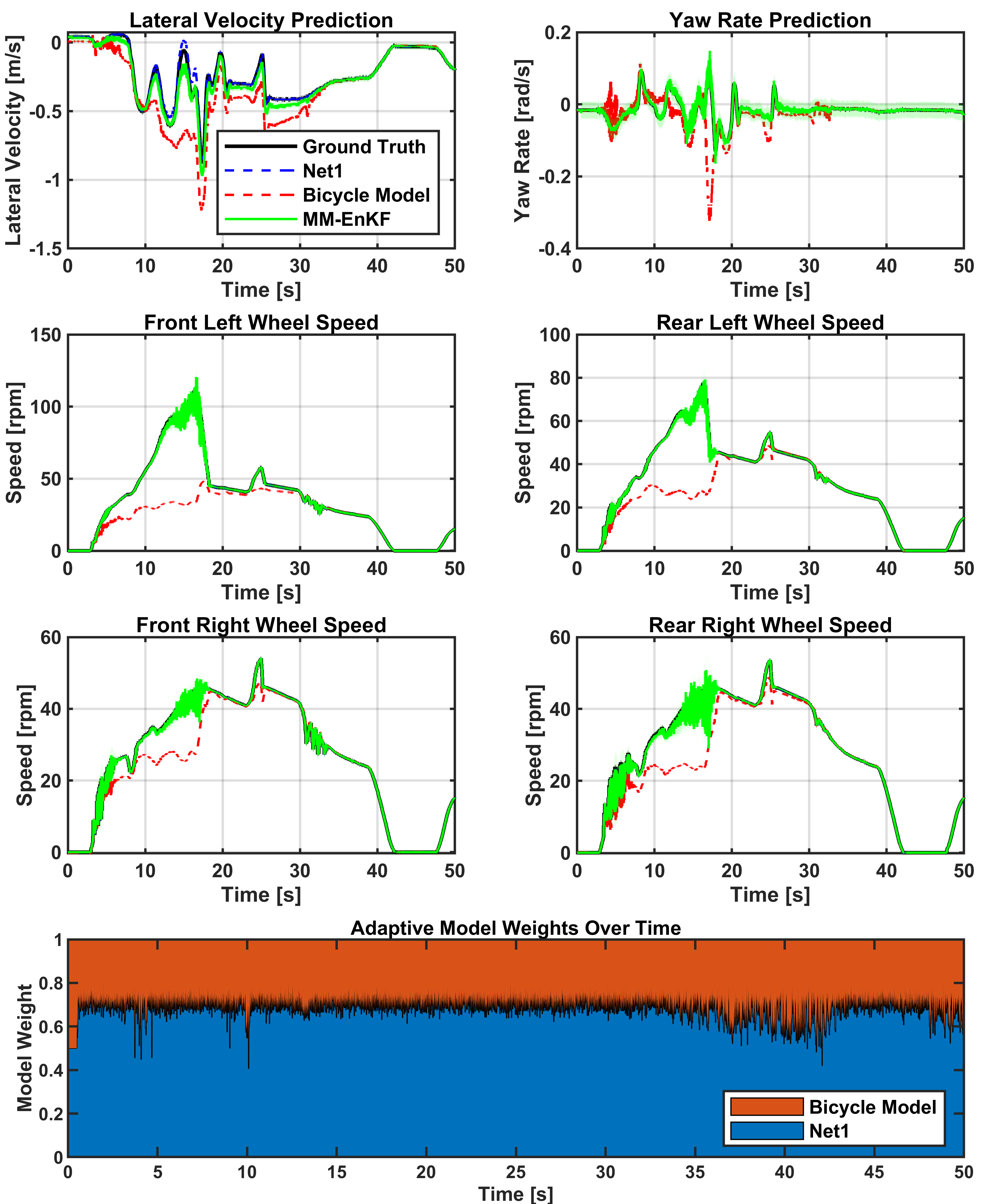}
    \caption{Performance of Method 2: Ensemble-Based Covariance Propagation.}
    \label{fig:MMeKF_results}
\end{figure}

Turning to our second methodology, the Ensemble-Based Covariance Propagation, we applied this approach to the same dataset to ensure statistical validity and consistent evaluation across methods. We maintained the first neural network trained in the previous approach and replaced the second network with a bicycle model representing the physical dynamics of the vehicle. The results and performance of this method are illustrated in Figure~\ref{fig:MMeKF_results}. Notably, during occasions with higher slip angles and nonlinear tire behavior, the bicycle model demonstrated limitations in capturing the complex vehicle dynamics. In these scenarios, our framework automatically assigned greater weight to the neural network, which demonstrated superior estimation performance. This automatic weighting adjustment highlights the adaptive nature of our consensus fusion approach, which dynamically balances the contributions of heterogeneous models based on their demonstrated reliability in current operating conditions.

Considering the results of both methodologies, it is evident that both approaches are validated and demonstrate excellent performance in vehicle state estimation. The consensus multi-model framework successfully integrates the complementary strengths of physics-based and data-driven models, providing robust estimation across diverse operating conditions. This is particularly valuable for critical vehicle states that cannot be directly measured but are essential for advanced vehicle control systems.

\section{Conclusion}

This paper presented a novel consensus multi-model Kalman filter framework that successfully bridges physics-based and data-driven models for robust vehicle state estimation. By developing two distinct methodologies—a Koopman operator-based linearization approach and an ensemble-based method—we effectively addressed the critical challenge of covariance propagation in heterogeneous model integration. The iterative consensus fusion procedure dynamically weighs different models based on their demonstrated reliability, enabling adaptive performance across diverse operating conditions.

Comprehensive experimental validation using an electric all-wheel-drive Equinox vehicle confirmed that our framework significantly outperforms single-model methods, with particularly pronounced advantages during challenging maneuvers and varying road conditions. The framework demonstrated exceptional capability in maintaining estimation accuracy during complex scenarios where traditional methodologies typically fail, including emergency maneuvers and low-friction surfaces.

The proposed methodology establishes a new paradigm for vehicle state estimation that leverages complementary strengths while mitigating individual weaknesses of different modeling approaches. This integration provides well-calibrated uncertainty quantification essential for safety-critical autonomous driving applications. 

\section*{Acknowledgements}
This document is the results of the research project funded by the Natural Sciences and Engineering Research Council of Canada (NSERC) under Grant ALLRP 566320 – 21.

\printcredits

\bibliographystyle{model1-num-names}

\bibliography{references}

\end{document}